\documentclass[12pt,epsfig]{article}
\usepackage{graphicx}
\def\lesim{ \;\raisebox{-.7ex}{$\stackrel{\textstyle <}{\sim}$}\; }
\def\be{\begin{equation}}
\def\ee{\end{equation}}
\def\slash#1{#1 \hskip-0.45em /}

\begin{document}
\begin{flushright}
IPPP/08/42 \\
DCPT/08/84 \\
9th June 2008 \\

\end{flushright}
%\begin{center}

%{\bf One-loop $gg\to b\bar b$ contribution - the main irreducible background to
%exclusive $H\to b \bar b$ production at the LHC)\\
\vspace*{3cm}

\begin{center}
{\Large \bf One-loop $gg\to b\bar b$ effects in the main \\[2mm]
irreducible background to exclusive $H\to b \bar b$ \\[4mm]
 production at the LHC }

\vspace*{1cm} \textsc{A.~G.~Shuvaev$^{a}$, V.A.~Khoze$^{a,b}$, A.~D.~Martin$^{b}$ and M.~G.~Ryskin$^{a,b}$}\\

\vspace*{0.5cm}$^a$ Petersburg Nuclear Physics Institute, Gatchina,
St.~Petersburg, 188300, Russia \\[0.5ex]
 $^b$ Department of Physics and Institute for
Particle Physics Phenomenology, \\
Durham University, DH1 3LE, UK \\%

%\end{center}

\vspace*{1cm}

\end{center}

{\bf Abstract:} We calculate the amplitude of $gg\to b\bar b$
production for the colour singlet,  $J_z=0$ di-gluon state at
${\cal O}(\alpha_S^2)$ order. We consider the cancellation, and a realistic
cutoff, of the infrared divergent terms. We show that the one-loop radiative QCD contributions effectively {\it reduce} the Born level result for the central exclusive $b\bar b$ cross section
at the LHC. This process is essentially the only irreducible QCD background to the exclusive $H\to b \bar b$ signal. \\
%in comparison
%with the central exclusive $b\bar b$ cross section, calculated in
%Born approximation, the one-loop radiative QCD correction
%strongly
%suppresses the irreducible $b\bar b$ background to the central
%exclusive diffractive Higgs boson production at the LHC.

%to the Born
%central exclusive $b\bar b$ production.

\vspace{1cm}
\section{Introduction}

\begin{figure}
\begin{center}
 \includegraphics[height=4cm]{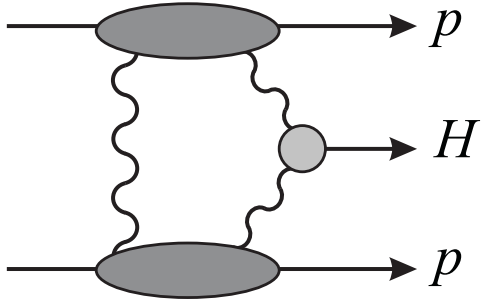}
\caption{\sf The mechanism for central exclusive Higgs production, $pp\to p \oplus H \oplus p$. The Higgs is produced by the fusion of two so-called active gluons. The screening gluon on the left is required to ensure a colour neutral system across the rapidity gaps. We do not show the additional screening corrections which must be included to ensure the survival of the rapidity gaps from population from secondaries resulting from soft rescattering.}
\label{fig:pHp}
\end{center}
\end{figure}

The search for, and  identification of,
the Higgs boson(s) is one of the main goals of the LHC. Once the Higgs
boson is discovered, it will be of primary interest  to determine its
spin and parity, and to measure precisely its mass and couplings, in
particular the $Hb\bar {b}$ Yukawa coupling.
% the coupling of Higgs boson to the $b$-quark.
%
An important contribution to the comprehensive study of the Higgs sector
can be provided by central exclusive diffractive (CED) Higgs boson production,
\be
pp\to p \oplus H \oplus p,
\ee
where the
$\oplus$ signs denote the presence of the large rapidity
gaps (LRG); see, for example \cite{KMR}-\cite{jf}. The process is sketched in Fig.~\ref{fig:pHp}. In such an exclusive process there is no hadronic
activity between the outgoing protons and the decay products of the
central (Higgs) system.
The  $b \bar b$ decay mode of the Higgs is especially attractive, since its observation would allow
a detailed study of the MSSM Higgs sector \cite{KKMRext}-\cite{clp}. Indeed, for certain BSM scenarios, it may
become {\it the} Higgs discovery channel \cite{KKMRext,fghpp}.
The experimental study of central exclusive Higgs production is
one of the key theoretical motivations behind the
FP420 project \cite{LOI,cox} which proposes to complement the
CMS and ATLAS experiments at the LHC
by installing  additional
forward proton detectors 420 m away from the interaction region.

To avoid the production of new secondaries
% by the colour flow
across the LRG,
the colour flow caused by the active gluons (which participate in
the $gg\to H$ fusion) must be screened by another, more soft,
gluon, see Fig.~\ref{fig:pHp}. Thus, in the exclusive process the central system
(the Higgs boson) is actually generated
by a specific $gg^{PP}$ gluon state, where the $PP$ superscript is to indicate that
that each hard (active) gluon
comes from colour-singlet $t$-channel (Pomeron) exchange.
% the hard gluons, which interact to form
%he central system, originate within the overall colourless (Pomeron)
%t-channel exchanges.}.
 Moreover, the presence of the second $t$-channel screening gluon leads to an additional loop integration.
The integration over the transverse momentum in this loop results
in a $J_z=0$, CP-even selection
rule~\cite{Liverpool, KMRJz}. Here $J_z$ is the projection of the total angular momentum
along the proton beam axis.

The $J_z=0$ selection rule is one
of the major reasons why central exclusive production is so
attractive for Higgs studies. First, it readily permits a clean determination
of the quantum numbers of the observed new resonance, which
 will be dominantly produced in a scalar state.
A second direct consequence of this rule is the unique possibility to study directly the $Hb\bar {b}$
 Yukawa coupling of
the Higgs-like bosons  in central exclusive processes. The potentially copious $b$-jet (QCD) background is controlled by
a combination of the $J_z=0$
selection rule \cite{Liverpool,KMRJz} (which
strongly suppresses the leading-order QCD $b \bar b$ production), the colour and spin factors and, finally, the
excellent resolution of the missing mass to the measured outgoing protons\footnote{Current studies \cite{LOI} show that the missing-mass
resolution $\sigma$ will be of the order 1\% for a 140~GeV Higgs
assuming that both protons are detected at 420 m from the interaction point. The equality of the accurate missing-mass measurement of $M_H$ with its mass determined from its decay products allows the background to be considerably suppressed.}.
%from the forward proton detectors.
It is the possibility to observe directly the dominant $b \bar b$ decay mode
of the SM Higgs with $M_H\lesim 140$~GeV that
first attracted attention to exclusive production at the LHC.
As is well known,
the direct determination of the $Hb\bar {b}$  coupling
appears to be very difficult for other search channels at the LHC.

We emphasize, that for forward going protons at the LHC,
 the Higgs signal is produced by gluons
in a $J_z=0$ state whereas the LO QCD backgrounds are primarily initiated by initial states with  $|J_z|=2$. The $J_z=0$ background contribution is suppressed for
large angles by a factor $m_b^2/E_T^2$, where $E_T$ is the
transverse energy of the $b$ and $\bar{b}$ jets,
 see for example \cite{FKM,krs}.
As discussed in \cite{BKSO,krs},
the physical origin of this suppression is
related to the symmetry properties of the Born helicity amplitudes
$M_{\lambda_1,
\lambda_2}^{\lambda_q,\lambda_{\bar q}}$ describing the binary background process
\begin{equation}
g(\lambda_1, p_A) \: + \: g (\lambda_2, p_B) \;
\rightarrow \; q
(\lambda_q, p_1) \: + \: \overline{q} (\lambda_{\bar q}, p_2) \/ .
\label{eq:a1}
\end{equation}
Here, the $\lambda_i$ label the helicities of the incoming gluons, and
$\lambda_q$ and $\lambda_{\bar q}$ are the (doubled) helicities of the
produced quark
and antiquark.  The  $p$'s denote the particle
four-momenta ($p_A^2=p_B^2=0$, $p_{1,2}^2=m^2$), with
$p_A+p_B=p_1+p_2$ and $s=(p_A+p_B)^2$.
It was shown in \cite{FKM} that for a colour-singlet,  $J_z = 0$, initial state,
$(\lambda_1=\lambda_2\equiv \lambda)$
the Born quark-helicity-conserving (QHC) amplitude
with $\lambda_{\bar q} = -\lambda_q$ vanishes\footnote{It is worth noting that in the massless limit
Eq.~(\ref{eq:a2}) holds for any colour state of initial gluons.
This is a consequence of the general property, that the non-zero massless tree-level
amplitudes should contain at least two positive or two negative helicity
states, see for example \cite{mhv1}. It is an example of the more general Maximally-Helicity-Violating
amplitude (MHV) rule, reviewed for example in \cite{MP}.}
\begin{equation}
 M_{\lambda, \lambda}^{\lambda_q, -\lambda_q} \; = \; 0.
\label{eq:a2}
\end{equation}
For the quark-helicity-non-conserving (QHNC) amplitude for
large angle
production we have
\begin{equation}
M_{\lambda, \lambda}^{\lambda_q, \lambda_q} \; \sim \; {\cal O} \left (
\frac{m_q}{\sqrt{s}}
\right ) \: M_{\lambda, - \lambda}^{\lambda_q, -\lambda_q},
\label{eq:a3}
\end{equation}
where the amplitude on the right-hand-side displays the dominant
helicity configuration of the LO background process.

%As we already mentioned, the $m_b^2/s$ suppression is especially critical in controlling the
%two-jet $b \bar b$ background.
The  main sources of background to the exclusive $H\to b \bar b$ production at the LHC
were discussed in detail in \cite{DKMOR,krs,hkrstw,insight}.
It was shown that all backgrounds are strongly  suppressed and controllable and,
in principle, can be
further reduced by the appropriate optimized cuts on the final state particle
configurations in such a way that the signal-to-background ratio $S/B$ is
of order 1
(or may be even better for MSSM Higgs \cite{KKMRext,hkrstw,clp}).

Among all the QCD backgrounds, the $m_b^2/E_T^2$-suppressed di-jet $ b \bar b$ production
is especially critical, since it is practically the only one irreducible
background source which cannot be decreased, either by improving the hardware
(as in the case, when the prolific
$gg^{PP}\to gg$ subprocess  mimics $ b \bar b$
 production, and the outgoing gluons are misidentified as
 $b$ and $\bar{b}$ jets \cite {DKMOR})
or, for example, by cuts on the three-jet event topology
(as in the case of large-angle gluon radiation in the process
$gg (J_z=0)\to q\bar q g$, discussed in \cite{krs}).
Therefore, the precise calculation of the QHNC  background
contribution becomes of primary importance for quantifying the signal-to-background ratio, and the accurate evaluation of the statistical
significance of the $H\to b \bar b$ signal.
 In \cite{DKMOR,hkrstw} it was estimated that the contribution from this source was $B/S \sim 0.4$, when the Born formula for the binary cross section was used. The same LO result for the background subprocess
 was used for the evaluation
of a statistical significance of the MSSM Higgs boson signal
in \cite{hkrstw,clp}. However, it is known (\cite{FKM,krs}) that higher-order QCD effects may strongly
affect the LO result.

{\it First}, there is a  reduction coming from
the self-energy insertions in the $b$-quark propagator, that is
from  running of $b$-quark mass from $\overline{m}_b
(\overline{m}_b)$  to its
value, $\overline{m}_b (M_H)< \overline{m}_b
(\overline{m}_b)$, at the Higgs scale.  Here
$\overline{m}_b (\mu)$
is the running $b$-quark mass in the $\overline{\rm MS}$ scheme
\cite{BBDM}.
%It is known that in the $H\to b
%\bar b$ decay width these single logarithmic (SL)
% ($ \alpha_S \ln \frac{M_H}{m_b}$) effects
%diminish the corresponding Born result by a factor of approximately two
%\cite{BL}.
%
%It is quite important that in central exclusive production a good
%signal-to-background ratio is achievable in the main Higgs decay
%($H\to b\bar b$) mode, where in conventional inclusive Higgs
%production the QCD background is huge. The point is that,
%specifically, for a $J_z=0$ initial state the Born quark helicity
%conserving (QHC) amplitude vanishes(see for example \cite{FKM})
%while the quark helicity non-conserving (QHNC) amplitude for large
%angle production is suppressed by the ratio,
%$m_q/E_T$, of the quark mass $m_q$ to the quark jet transverse energy
%$E_T$. This $m^2_b/E^2_T$ suppression of the QCD background cross
%section is critical in controlling the $b\bar b$ background.\\
%
%However the lowest $\alpha_S$ order Born cross section is strongly
%affected by the higher order corrections. First there is a reduction
%coming from the self-energy insertion into the $b$-quark propagator,
%that is from the running of the $b$-quark mass from ${\overline
%m}_b(m_b)$ to its value ${\overline m}_b(M_H)<
%{\overline m}_b(m_b)$ at the Higgs scale. Here
%${\overline m}_b(\mu)$ is the running mass in the $\overline{MS}$
%scheme.
{\it Secondly}, in our case, where $m_b \ll M_H$, double-logarithmic corrections of
the form $(\alpha_S/\pi)\ln^2(M_H/m_b)$ are potentially very important.
These are  related to the so-called non-Sudakov form factor, $F_q$, in the
cross section which arises from the virtual diagrams with
gluon exchange,
see~\cite{FKM,JT,MS,MSK}.\\
For the case of the  $\gamma\gamma (J_z=0)\to b\bar b$
process the complete one-loop result was first calculated in
\cite{JT}. For the photon-photon reaction,
the double-logarithmic (DL) asymptotics for $F_q$
%( the $J_z=0$ QHNC case)
 has the form
\begin{equation}
F_q(L_m)\ =\ \sum_n c_n\left(\frac{\alpha_S}{\pi}L^2_m \right)^n
\label{eq:FLm}
\end{equation}
with $L_m \equiv \ln(M_H/m_b)$, $c_0=1$ and $c_1=-8$~\cite{FKM,JT}.
The second
(negative) term in (\ref{eq:FLm}) is anomalously large, and dominates the
Born term for $M_H>100$ GeV. Clearly, this dominance undermines the results
of any analysis based on the Born approximation\footnote{For $gg^{PP}$ fusion, rather than $\gamma\gamma$ fusion, we might, a priori, anticipate that colour factors could make the coefficient $c_1$ even larger.}.
The physical origin of this non-Sudakov form factor was elucidated
in~\cite{FKM}, where the explicit DL calculation at the two-loop level
was performed. It was also shown that the two-loop calculation should be sufficient for a reliable evaluation of the DL effects. This
was confirmed by a more comprehensive all-orders study~\cite{MS,MSK}.

Recall that in the photon-photon case, the two-loop expression for the
$F_q$ is \cite{FKM}
\begin{equation}
F_q(L_m)\ =\ (1-3{\cal F})^2+\frac{{\cal F}^2}3
\left(1+\frac{C_A}{C_F}\right)\ ,
\end{equation}
with
\begin{equation}
{\cal F}\ =\ \frac{\alpha_S}{\pi}C_FL^2_m,
\end{equation}
where $C_F=(N^2_c-1)/2N_c$, $C_A=N_c$ and $N_c=3$ is the number of
colours.
The non-Sudakov logarithms come from kinematical regions
%diagrams
 where one of the
{\it quark} propagators is soft. As well known, there are other DL
effects (the well known Sudakov
logarithms~\cite{Sud}) which arise from virtual soft {\it gluon}
exchange. As discussed in~\cite{FKM}, in the case of quasi-two-jet
configurations, the Sudakov and non-Sudakov effects can be factorized to good
accuracy, since they correspond to very different
virtualities of the internal quark and gluon lines. For final
state radiation, the Sudakov effects can be implemented in parton
shower Monte Carlo models in the standard way. For the $gg^{PP}$ initial
state, the Sudakov factors are explicitly incorporated in the
unintegrated gluon densities, see~\cite{KMR,KMRpr}.

Unfortunately, from a phenomenological perspective, it is
dangerous to rely on the DL results, since experience shows that
formally subleading (SL) corrections may be numerically important.
That is why an accurate evaluation of
the QCD $b \bar b$ background to the central exclusive process
\be
pp\to p\oplus(H\to b\bar b)\oplus p
\ee
 requires, first of all, the
calculation of
the exact one-loop correction to the Born
$gg^{PP}\to b\bar b$ amplitude, that is to the hard
subprocess $gg\to b\bar b$ in a colour-singlet $J_z=0$ initial
state.

The calculation is described in Section 2, where the result is
presented as the sum of  standard integrals corresponding to the
box, triangle and self-energy Feynman diagrams with the scalar
propagators. In Section 3 the infrared divergencies of the virtual loop
amplitude are discussed. These divergences are cancelled either by
real gluon emission or by the diagrams where an additional virtual
gluon is emitted off the second $t$-channel (screening) gluon in the
whole $pp\to p\oplus b\bar b\oplus p$ amplitude. Numerical
estimates are illustrated in Section 4.

\section{Calculation of  the $gg\to b\bar b$ amplitude}

%We will throughout denote the incoming gluons' momenta as
%$p_A$ and $p_B$, $p_A^2=p_B^2=0$
%and quark and antiquark ones as $p_{1,2}$, $p_{1,2}^2=m^2$,
%p_A+p_B=p_1+p_2$.

Here, and in what follows, we consider the amplitude, defined as
\begin{equation}
T_{gg\to b\bar b}=\sum_{a,b}\sum_{\epsilon_1,\epsilon_2}
M_{ab}^{\epsilon_1,\epsilon_2}\delta_{ab}\delta_{\epsilon_1\epsilon_2},
\end{equation}
where $a$ and $b$ are the colour indices ($a,b=1,2,...,N_c^2-1$)
and $\epsilon_{1,2}$ are the transverse polarization vectors of the
incoming gluons. Introducing the outgoing quark spinors, the amplitude can
be written as
\begin{equation}
T_{gg\to b\bar b}\,=\,\overline u(p_1)T(\slash p)v(p_2),
\end{equation}
 where the matrix $T$ is built from the 4-momenta $p$ involved in the reaction, $\slash p \equiv p_\mu
\gamma^\mu$.  There are three independent 4-momenta, which we take to be $p_1$,
$p_2$ of $b$ and $\bar b$, and the difference of the incoming gluon momenta, $p_A-p_B$. The matrix $\slash p_1$ can be moved to the left,
and $\slash p_2$ to the right, until they disappear upon acting on the spinor,
\be
\overline u(p_1)\slash p_1=m\overline u(p_1),~~~~~~~\slash p_2 v(p_2)=-mv(p_2),
\ee
where $m \equiv m_b$.
This rearrangement reduces the amplitude to two spinor structures, corresponding to
the following helicity-violating and helicity-conserving terms\footnote{In principle, there are two other possible structures
related to the fourth vector $e_\mu \sim
\varepsilon_{\mu\nu\lambda\sigma}p_1^\nu(p_A-p_B)^\lambda p_2^\sigma$
namely, $u(p_1)\gamma^5v(p_2)$ and
$\overline u(p_1)(\slash p_A-\slash p_B)\gamma^5 v(p_2)$,
but they do not appear for our subprocess.}
\begin{equation}
T_{gg\to b\bar b}\,=\,\overline u(p_1)v(p_2)\,T_1\,
+\,\overline u(p_1)(\slash p_A-\slash p_B)v(p_2)\,T_2~.
\label{eq:T1T2}
\end{equation}
The scalar coefficients $T_{1,2}$ depend on the invariants
\be
s=(p_1+p_2)^2,~~~~~t=(p_A-p_1)^2~~~~~u=(p_A-p_2)^2.
\ee
The function
$T_1$ is symmetric with respect to the $t$-$u$ interchange, while $T_2$ is antisymmetric.
Their calculation requires an evaluation of the spinor traces, which
can easily be carried out with the help of an analytical program.

\subsection{The Born contribution}
As discussed above,
only the helicity-violating piece
 contributes at the Born level to the $J_z=0$ amplitude, that is the sum $(++)\ +\ (--)$ of the helicities of the incoming gluons,
\begin{equation}
T_1^{\rm Born}\,=\,4\pi\alpha_S\,C_F\,2m\left[\frac{1}{m^2-t}
+\frac{1}{m^2-u}\right],\qquad
T_2^{\rm Born}\,=\,0.
\end{equation}
The loop integration is reduced to master scalar integrals,
and the final result is written in terms of the set of these integrals.
%through the set of which the final result is expressed.
The Feynman gauge is used  in the
$\overline{\rm MS}$-scheme with the dimensional parameter of order
of
%% incoming gluon virtuality
the characteristic virtualities of the process,
% relevant for the process,
$4\pi\mu^2=-k^2=s/4$.
    The ultraviolet (UV) divergencies are renormalized, to first
order in $\alpha_S$, by the gluon and quark $Z$-factors, mass and charge
renormalization in the Born term.

For central exclusive
production, the effective luminosity of the incoming active gluons,
\begin{equation}
\frac{M^2\partial{\cal L}(gg^{PP})}{\partial y\partial M^2}\,=\, \hat
S^2 L^{\rm excl},
\label{eq:lump}
\end{equation}
is given by the integral expression \cite{KMRpr}
\begin{equation}
L^{\rm excl}\,=\,\left(\frac \pi{(N^2_c-1)b}\int\frac{dQ^2_t}{Q^4_t}
f_g(x^+_1,x'^+_1,Q^2_t,\mu^2)f_g(x^-_2,x'^-_2,Q^2_t,\mu^2)\right)^2,
\label{eq:lum}
\end{equation}
where the integration is over the transverse momentum of the gluon loop in Fig.~\ref{fig:pHp}.
Here $b$ is the $t$-slope, corresponding to the momentum transfer
distributions of the colliding protons, and $x_{1,2}$ are the light-cone
momentum fractions carried by the active gluons\footnote{
The superscript $+$  ($-$) is to indicate the light-cone momentum
fraction of the momentum $q_1,~(q_2)$ of the first (second) colliding proton, that is $Q_\mu=x'^+_1q_{1\mu}+x'^-_2q_{2\mu}+Q_{t\mu}$.}.
 The colour flow due to the active gluons is screened by the second $t$-channel
gluon, which carries the momentum fractions $x'^+_1$ and $x'^-_2$. The gap
survival factor $\hat S^2$ accounts for the soft rescattering effect, that is $\hat S^2$ is the probability that the rapidity gaps are not populated by secondaries produced in possible soft rescattering, see \cite{KMR,KMRpr,soft}.
The integral over the gluon transverse momentum $Q_t$ is convergent, in
both the ultraviolet and infrared regions. The conribution of the low $Q_t$
domain is suppressed by Sudakov-like form factors
incorporated in the unintegrated gluon densities $f_g$.

Due to the presence of the second $t$-channel (screening)
gluon, there is no infrared divergency
in the matrix element of the `hard' $gg(J_z=0)\to b\bar b$ subprocess.
This opens up the possibility to regularize  the infrared (IR)
divergences by
% the introduction of
 a fictitious gluon mass $m_g$.
 In order to preserve gauge invariance, the mass $m_g$ can be formally introduced
 via the Higgs mechanism\footnote{This way of introducing the
  infrared cutoff $m_g$ was used, for example,
   in the original BFKL calculations, see \cite{klf}.}.
 Of course, this will generate additional diagrams
 with Higgs boson exchange or production. Such  Higgs boson exchange will provide `cross-talk' between the `hard' matrix element and the
 second $t$-channel (screening) gluon. As a result, we set $m_g=Q_t$, that is,
introduce a physical infrared cutoff in the exclusive $b\bar b$
production. After this, the contribution of the ``artifical'' Higgs boson diagrams can
be eliminated by choosing the mass of the ``Higgs'' boson to be
very large, $M^2_H \gg s$. Since there is no need to regularize the integral (\ref{eq:lum}),
here we will work in the $D=4$ world, so the polarizations of our gluons are
constrained to $D=4$ space.

\subsection{One-loop effects}

\begin{figure}
 \begin{center}
 \includegraphics[height=10cm]{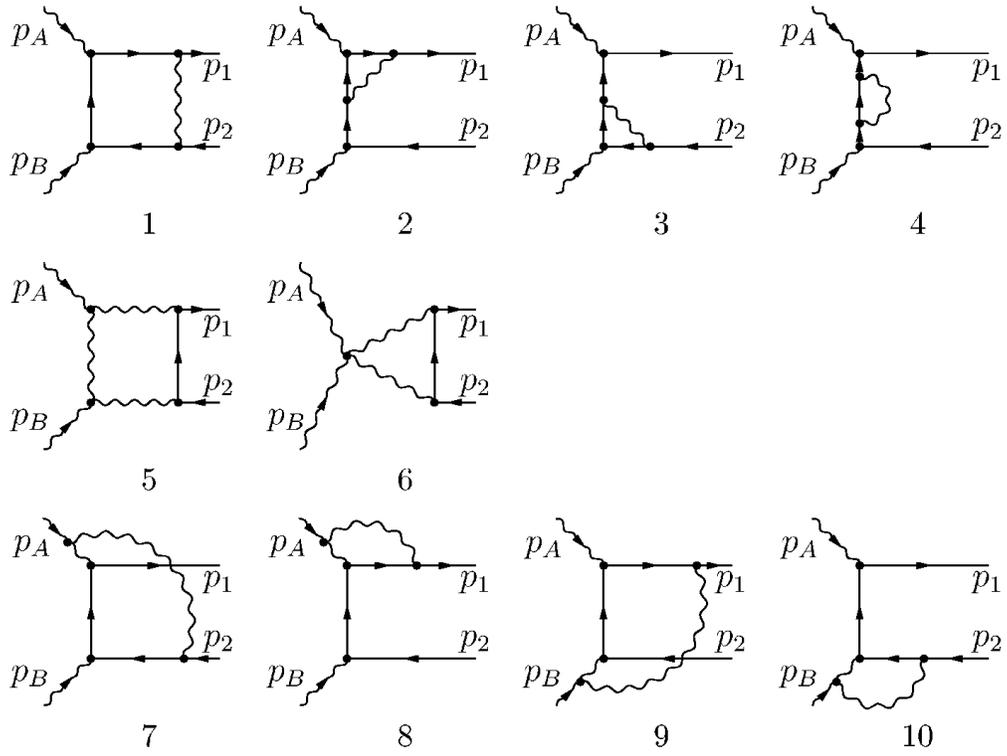}
 \caption{\sf One-loop diagrams contributing
 to the $gg\to b\bar b$ process }
 \label{fig:ggqq}
\end{center}
 \end{figure}

Both the structures $T_{1,2}$ in (\ref{eq:T1T2}) can contribute to the one-loop correction. We
begin with the helicity-conserving part
$$
T_2(s,t,u)\,=\,A_2(s,t,u)-A_2(s,u,t).
$$
We retain only the quark mass $m=0$ contribution. The function $A_2(s,t,u)$ is given by the set of Feynman diagrams
shown in Fig.~\ref{fig:ggqq}. The second term in the expression for $T_2(s,t,u)$
takes care of $t \leftrightarrow u$ interchange.
When performed naively, a straightforward calculation results in a complete cancellation between the
two terms. So we have $T_2(s,t,u)=0$ at the one-loop level,
similar to the Born term\footnote{At first sight, it appears that the zero value of $T_2$ is in
contradiction with  \cite{adr,ng}, which lead to a non-zero result for the one-loop
(cut-non-reconstructible) QHC amplitude $gg(J_z=0)\to b \bar b $
in dimensional regularization, assuming that the incomimg gluons
are on-mass-shell. This non-zero cut-non-reconstructible contribution
comes from a ratio of the form $\epsilon/\epsilon=const$, where the
denominator (that is, the factor $1/\epsilon$) is of infrared origin.
However, in our case, due to the presence
of the screening gluon, the (transverse) size of the interaction region is
limited by a value of order of $\sim 1/Q_t$. This generates a dynamical
infrared cut-off at a scale $\sim Q_t^2$, which, in our
calculation, is taken care of by introducing an effective gluon mass $m_g=Q_t$.
As a result, the infrared singularities are absorbed in
the unintegrated gluon structure functions.
The factor $1/\epsilon$ is replaced by $\ln(\mu/m_g)$, and as
$\epsilon\to 0$ we obtain $T_2=0$. Thus
our procedure is quite different from the on-mass-shell calculations.
%---- %The %non-zero amplitude $T_2$ in \cite{adr,ng} was obtained for
%$gg(J_z=0)\to b \bar b $ %process at ${\cal O}(\alpha_S$) order  after
%dimensional regularisation. There the non-zero cut-nonreconstructible
%contribution comes from a ratio of the form $\epsilon/\epsilon=const$,
%where the $\epsilon$ in the numerator corresponds to the additional
%polarisations of the gluon (lying in %$\epsilon$ space)  while the
%denominator %(i.e.  the factor $1/\epsilon$) is of infrared origin.
%There is no such contribution %in our case since the size of the
%interaction region %($\sim 1/Q_t$) is limited by the presence of the
%second, screening, %$t$-channel gluon
%(with momentum fractions $x'_1,\ x'_2$) and, therefore,
%all  integrals have no IR divergency.}.
%-------
}.

The QHNC part of the amplitude is given by the
two terms,
\be
T_1(s,t,u)\,=\,A_1(s,t,u)\,+\,A_1(s,u,t),
\ee
where the second term, as in the previous case, comes from the $u$-channel crossing.
Recall that the amplitude $T_1$ is symmetric with respect to the $t$-$u$
interchange. As in the Born term, the one-loop
amplitude vanishes if we set $m=0$.   Here we present an expression for the amplitude to
lowest order in mass, that is
\be
A_1\,=\,4\pi\alpha_S\,\frac
{\alpha_S}{4\pi}\,m\, \bigl[A_0\,+\,A_q\,+\,A_g\,+\,A_{qg}\bigr],
\label{eq:A1T1}
\ee
where $m$ is the quark pole mass, $m=\overline{m}_b(\overline{m}_b)$, and the running
coupling constant is taken at the scale $\mu$, $\alpha_S=\alpha_S(\mu^2)$. The function $A_0$ is made up of pieces coming from the logarithmically divergent
scalar master integrals with two propagators. The $1/\epsilon$ terms,
appearing in dimensional regularization, are subtracted by the counterterms coming
from the field $Z$-factors, quark mass and coupling constant renormalization
in the Born term. To be precise, $A_0$ collects what is not
included in the functions $I$, which result from the master integrals. It depends on
which form has been assumed for the gluon polarizations in dimensional regularization. That is, whether the transverse tensor lies exactly in two-dimensional space $D-2=2$, or whether it has an admixture
of the extra dimension, $D-2=2+2\epsilon$.
As we  discussed, in central diffractive  production the active incoming gluons are screened
 (in the whole amplitude of {the \it exclusive} process) by a
 second $t$-channel gluon, see Fig.~\ref{fig:pHp}.
 % which was needed to compensate the colour flow  across the gap.
  In this external loop we have neither infrared
 nor UV divergencies. Thus the polarisations of incoming gluons lie in
 two-dimensional  $D-2=2$ space. Moreover, when using the unintegrated gluons
 obtained via the KMR prescription \cite{KMRMR}, from the integrated
 $\overline{\rm MS}$ gluon given by the global parton analyses, we have to account
 for the fact that the whole gluon renormalisation factor $Z_3$ is
 already included in the incoming parton distribution. So the matrix element
 of the `hard' subprocess must be calculated with the $Z$ factors equal to 1
 for all `external' lines (just as is the case when the external lines
 are on-mass-shell). In other words, here we present the result of the
 calculation\footnote{The renormalisation factor, $Z_2$, of the $b$-quark
 is included in the evolution which describes the fragmentation of the $b$-quark jet.} with $Z_3=Z_2=1$.

The result obtained for $Z_2=1$, $Z_3=1$, with the infinite parts subtracted, has the form,
\be
A_0^{D-2=2}\,=\,\frac 1t\left[-4C_FN_c\ln\frac{m^2}{m_g^2}\,
+\,C_F\,\frac{5N_c^2-1}{N_c}\ln\frac{m^2}{4\pi\mu^2}\,-\,24C_F^2
\right].
\ee
The other functions in (\ref{eq:A1T1}) are given by the finite parts of the individual Feynman diagrams
shown in Fig.~\ref{fig:ggqq} with the colour coefficients $c_i$,

\begin{eqnarray}
c_1\,=\,c_4\,&=&\,C_F^2\,=\,\left[\frac{N_c^2-1}{2N_c}\right]^2 \nonumber \\
c_2\,=\,c_3\,&=&\,-\frac1{2N_c}C_F\, =\, -\frac1{2N_c}\frac{N_c^2-1}{2N_c}
\nonumber \\
c_5\,& =&\,N_c\,C_F\, =\,N_c\frac{N_c^2-1}{2N_c}
\nonumber \\
c_6\,& =&\,N_c^2-1\nonumber \\
\nonumber
c_7\,=\,c_9\,& =&\, -\frac12 N_c\,C_F\, =\, -\frac12 N_c\frac{N_c^2-1}{2N_c}
\nonumber \\
c_8\,=\,c_{10}\,& =&\, -\frac12 N_c\,C_F\, =\,
-\frac12 N_c\frac{N_c^2-1}{2N_c}.
\nonumber
\end{eqnarray}
The master integrals, $I^q_{0000}(s,t)$ etc, appearing in the expressions
below, are given in the Appendix A. We have
\begin{eqnarray}
A_q\,&=& \,4\,c_1I^q_{0000}(s,t)s
+\,I^q_{0101}(t)\frac{2[(2s + t)(c_2+c_3) + 2(2s - t)\,c_1  - 2\,c_4s]}
{s\,t},\nonumber \\
&& \nonumber \\
&& \nonumber \\
A_g\,&=&\, 4I^g_{0000}(s,t)c_5s
+\,3I^g_{1000}(s)[c_6 - 2c_5]
-\,4\,c_5I^g_{0100}(t) \nonumber \\
&&+\,I^g_{1010}(s)\frac{4[-(3s + 2t)c_5 + c_6s]}{s^2}
+\,I^g_{0101}(t)\frac{4(2s + t)c_5}{s\,t}, \nonumber \\
&& \nonumber \\
&& \nonumber \\
A_{qg}\,&=&\,I^{qg}_{0000}(u,t)\frac{ - [2s^2 + 3st + 3t^2](c_7+c_9)}{s}
\,+\,I^{qg}_{1000}(u)\frac{3(s + t)(c_7+c_9)}{s} \nonumber \\
&&+I^{qg}_{1000}(t)\frac{ - 3\,(c_7+c_9)\,t}{s}
\,+ I^{qg}_{0010}(u)\frac{\,(c_7+c_9)(2s+3t)}{s} \nonumber \\
&&-\,I^{qg}_{0010}(t)\frac{\,(c_7+c_9)(s+3t)+2(c_8+c_{10})s\,}{s}
\,+\,I^{qg}_{1010}(u)\frac{ - 4(c_7+c_9)}{s + t} \nonumber \\
&&+\,2I^{qg}_{1010}(t)\frac{[(c_8+c_{10})(t-2s) + 2(c_7+c_9)s]}{st}.
\nonumber
\end{eqnarray}

\section{The infrared contribution}
The amplitude $T_1$ contains logarithmic infrared (IR) divergences
which in the formulae presented above are regularized
by  the effective gluon mass $m_g$ cut-off.
%cutted off in our formulae by the effective gluon mass $m_g$.
These IR divergent terms are essentially the usual Sudakov form
factors, that is the probability not to emit  additional gluons in the
{\it exclusive} process $pp\to p\oplus b\bar b\oplus p$.

The Sudakov-like form factor, due to emission from the initial active gluons, is equal to
\begin{equation}
S(Q_t,\mu)=exp\left(-\int^{\mu^2}_{Q^2_t}\frac{\alpha_S}{2\pi}
\frac{dk^2_t}{k^2_t}\int_0^{1-\Delta}zP_{gg}(z)dz\right)
\label{sudg}
\end{equation}
with $P_{gg}(z)$ being the gluon-gluon Altarelli-Parisi LO splitting
function corresponding to real gluon emission,
and $\Delta=k_t/(\mu+k_t)$. It was already included in the
effective gluon-gluon luminosity ${\cal L}(gg^{PP})$ used to calculate
the exclusive cross section \cite{KMRpr}. Therefore, we have to subtract the
term
\begin{equation}
T^{\rm Born}_1\left(-\int^{\mu^2}_{Q^2_t}\frac{\alpha_S}{2\pi}
\frac{dk^2_t}{k^2_t}\int_0^{1-\Delta}zP_{gg}(z)dz\right)
\label{eq:sg}
\end{equation}
from the amplitude $T$.

As discussed above, because of the presence of a screening gluon,
the infrared cut-off is given by the
transverse momentum (virtuality) of the incoming active gluon,
and in the expressions for the amplitudes $A_g$ and $A_{qg}$,
where the `internal' gluon is radiated from a gluon line,
we replace $m_g$ by $Q_t$.
%Since the physical infrared cut-off is
%provided by the presence of the screening gluon, that is
%the infrared cut-off is given by the
%transverse momentum (virtuality) of the incoming active gluon,
%we put in the expressions for the amplitudes $A_g$ and $A_{qg}$,
%where the 'internal' gluon was emitted off the gluon line,
This cancels the main ($\propto\ln^2(m^2_g)$) part
of the IR divergency
\be
 T_1\simeq 2m(4\pi\alpha_S)N_cC_F\frac{\alpha_S}{\pi}\left(\frac{1}{t}+\frac{1}{u}\right)\ln^2(s/m_g^2).
\ee
The logarithmic IR divergency in the amplitude $A_q$ is
cancelled after accounting for the real soft gluon emission in the $b$-quark
jet. Usually, in Monte Carlo simulations and/or jet searching algorithms, such
emission is described by LO quark evolution. So we have to
subtract the term
\begin{equation}
T^{\rm Born}_1\left(-\int^{\mu^2}_{k^2_{t0}}\frac{\alpha_S}{2\pi}
\frac{dk^2_t}{k^2_t}\int_0^{1-\Delta_q}P_{qq}(z)dz\right),
\label{eq:sq}
\end{equation}
where now $P_{qq}(z)$ is the quark-quark splitting function;
the lower limit $k_{t0}$ is fixed by the experimental conditions -- gluons with transerse momenta with respect to $b$-jet axis
$k_t<k_{t0}$ are included in the definition of the jet. The kinematic
limit is $\Delta_q=2k_{t0}/\sqrt s$.
As a result, we put $m_g=k_{t0}$ in the expression\footnote{There still remains a contribution proportional to
the first power of $\ln(m_g)$, which is not cancelled by the subtractions
(\ref{eq:sg},\ref{eq:sq}). This contribution arises from large-angle
soft-gluon emission, when we cannot neglect the interference between the
emission from the gluon and from the quark lines. Such  interference,
hidden in the amplitude $A_{qg}$, is not included, either in the
definition of the jet or in the effective $gg^{PP}$ luminosity. The
corresponding IR divergency is cut off by the presence of the
screening gluon in the effective $gg^{PP}$ luminosity, that is by the
gluon transverse momentum $Q_t$; to mimic this fact we set
$m_g=Q_t$, as before.} for $A_q$.

\section{The double-logarithmic contributions}
As was discussed in the Introduction,  large double-logarithmic terms
can be of Sudakov or non-Sudakov origin. The Sudakov contributions
reflect the possibility to emit additional soft gluons. They are
absorbed (and subtracted) in the definitions of the $gg^{PP}$ luminosity
and in the prescription for the quark jet search. The non-Sudakov
logarithms come from the kinematical domain in which one of the quark propagators
in the diagram is soft. In the case of the $\gamma\gamma\to b\bar b$
process this contribution was numerically quite large
\be
T_1^{\rm non-Sud}(\gamma\gamma\to b\bar b)~\simeq~ T_1^{\rm Born}\cdot 3C_F\frac{\alpha_S}{4\pi}\ln^2(s/m^2_b),
\ee
see (\ref{eq:FLm}).  In our case, with a larger number of diagrams and
larger colour coefficients in the one-loop $gg\to b\bar b$ Feynman
graphs, there is a danger that we could find an even  larger non-Sudakov DL correction.
However, the situation appears to be different.
%is not the fact.
 Contributions which correspond to
diagrams with 3 gluons in the loop and to diagrams with 2 gluons have
different signs. This is analogous to the destructive interference between
the emission of a photon from the incoming and the outgoing electron for small angle scattering.
Unlike the $\gamma\gamma\to b\bar b$ case, here we have additional
double-logarithm contributions coming from $A_g,\ A_{qg}$, such that the
final result does not contain a large numerical coefficient
\be
T_1^{\rm non-Sud}(gg\to b\bar b)~\simeq~ T_1^{\rm Born}\cdot
(3C_F-N_c)\frac{\alpha_S}{4\pi}
\ln^2(s/m^2_b).
\ee
We see that, instead of the naively expected factor $3N_c$ (which
indeed comes from the first term of $A_g$, that is from the integral
$I^g_{0000}$), the coefficient in the sum of the $A_g$ and $A_{qg}$
amplitudes in front of the non-Sudakov
double-logarithm
%(in the sum of the $A_g$ and $A_{qg}$ amplitudes)
is proportional to $3N_c-4N_c=-N_c$.

\section{Discussion}

\begin{figure}
%[t]
 \begin{center}
 \includegraphics[height=10cm]{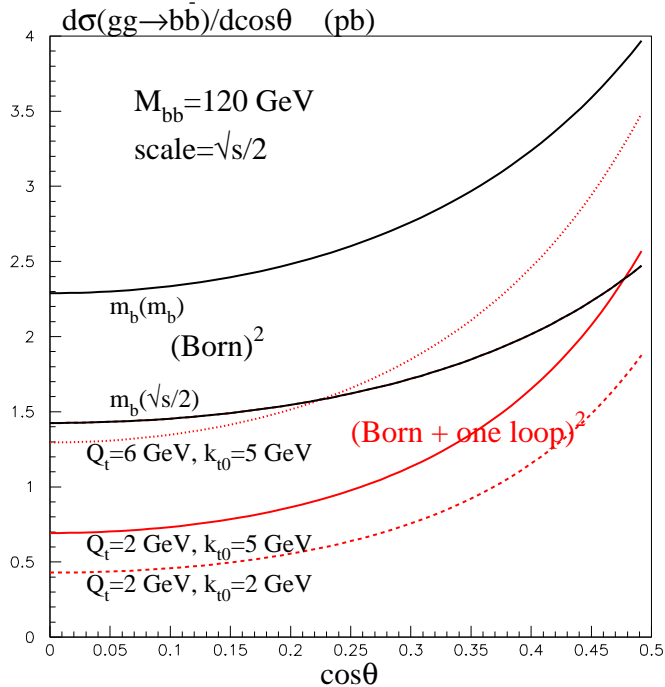}
\caption{\sf Angular dependence of the exclusive $b\bar b$ cross section for different choices of the infrared cutoffs.  For comparison, we also show the Born result for two choices of the running $b$-quark mass.}
\label{fig:ggbc}
\end{center}
\end{figure}

To evaluate the role of the one-loop correction numerically, we first calculate the cross section
\begin{equation}
\frac{d\sigma^{(0)+(1)}}{d\cos\theta}\,=\,\frac 1{32\pi s}\left(\frac
1{2(N_c^2-1)}(T^{\rm Born}_1+T_1)\right)^2 \,,
\label{eq:cros}
\end{equation}
that must be multiplied by the
effective gluon-gluon luminosity (\ref{eq:lump}) \cite{KMRpr}.
The results are shown in Fig.~\ref{fig:ggbc} for different
values of the infrared cutoffs $Q_t$ and $k_{t0}$. The scale is taken to be
$4\pi\mu^2=s/4$.

The value $Q_t=2$ GeV corresponds to the maximum of the integrand
in (\ref{eq:lum}) for the exclusive production of a Higgs boson of mass $M_H=120$ GeV at the LHC. The choice $k_{t0}=5$ GeV appears to be reasonable for the standard $b$-quark-jet searching; the gluons with transverse momentum (with respect to $b$-jet axis) of $k_t>5$ GeV can be
considered as separate jets. To demonstrate the dependence
of the cross section $\sigma^{(0)+(1)}$ on the values of $Q_t$ and $k_{t0}$
we also show the predictions for $Q_t=6$ GeV and $k_{t0}=2$ GeV.
We use a $b$-quark pole mass $\overline{m}_b(\overline{m}_b)=4.2$ GeV and take
$\alpha_S(M_Z)=0.118$.
For lower infrared cutoffs the probability not to emit an additional gluon decreases, and the cross section is smaller.

For comparison, we also show in Fig.~\ref{fig:ggbc} cross sections calculated
in the Born approximation with the same renormalisation scale
($s/4$) for the QCD $\alpha_S$ coupling and the $b$-quark pole mass
${\overline m_b}$ taken at the same scale ${\overline m_b}(\overline{m}_b)$ (upper curve). In addition we plot the Born result for a $b$-quark mass taken at the scale $\sqrt{s}/2$ (lower curve); this shows that a large part of the one-loop suppression of the cross section comes from the running of the $b$-quark mass.  However, note that for large-angle
scattering we observe a
stronger suppression of the cross section due to other radiative
corrections.

\begin{figure}
%[t]
 \begin{center}
 \includegraphics[height=10cm]{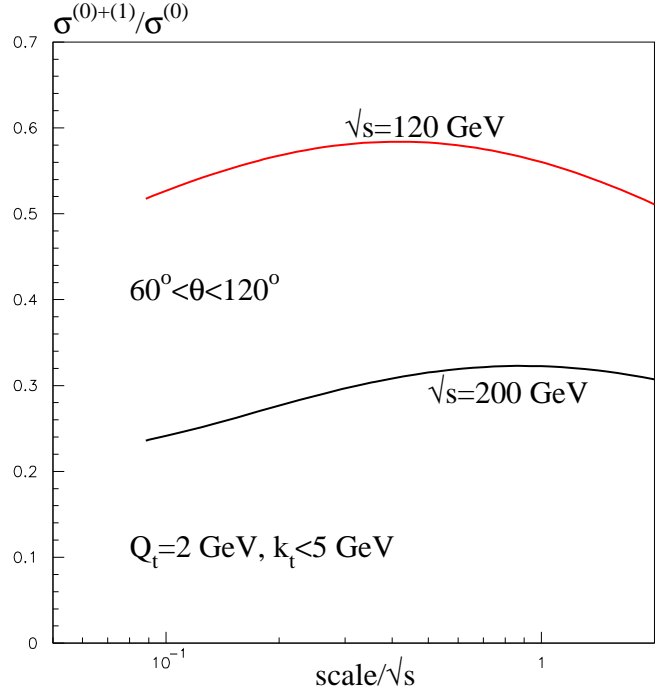}
 \caption{\sf The scale dependence of the ratio of the NLO exclusive
 $b\bar b$ cross section to that calculated in Born approximation.}
 \label{fig:ggbs}
\end{center}
\end{figure}
In Fig.~\ref{fig:ggbs}  we show the scale dependence of the ratio
$\sigma^{(0)+(1)}/\sigma^{(0)}$
of the whole one-loop cross section, integrated over the region
$60^o<\theta<120^o$ (that is $|\cos\theta|<1/2$),
to that calculated in Born
approximation with the pole mass ${\overline m_b(\overline{m}_b)}$ of $b$-quark. The result is
shown for two different masses of $b\bar b$ system, namely $M_{bb}=\sqrt s =120$
and 200 GeV. Here we put $Q_t=2$ GeV and $k_{t0}=5$ GeV. It is seen
that the scale dependence in the region of $4\pi\mu^2\sim s/4$ is
rather flat.

\begin{figure}
%[t]
 \begin{center}
 \includegraphics[height=10cm]{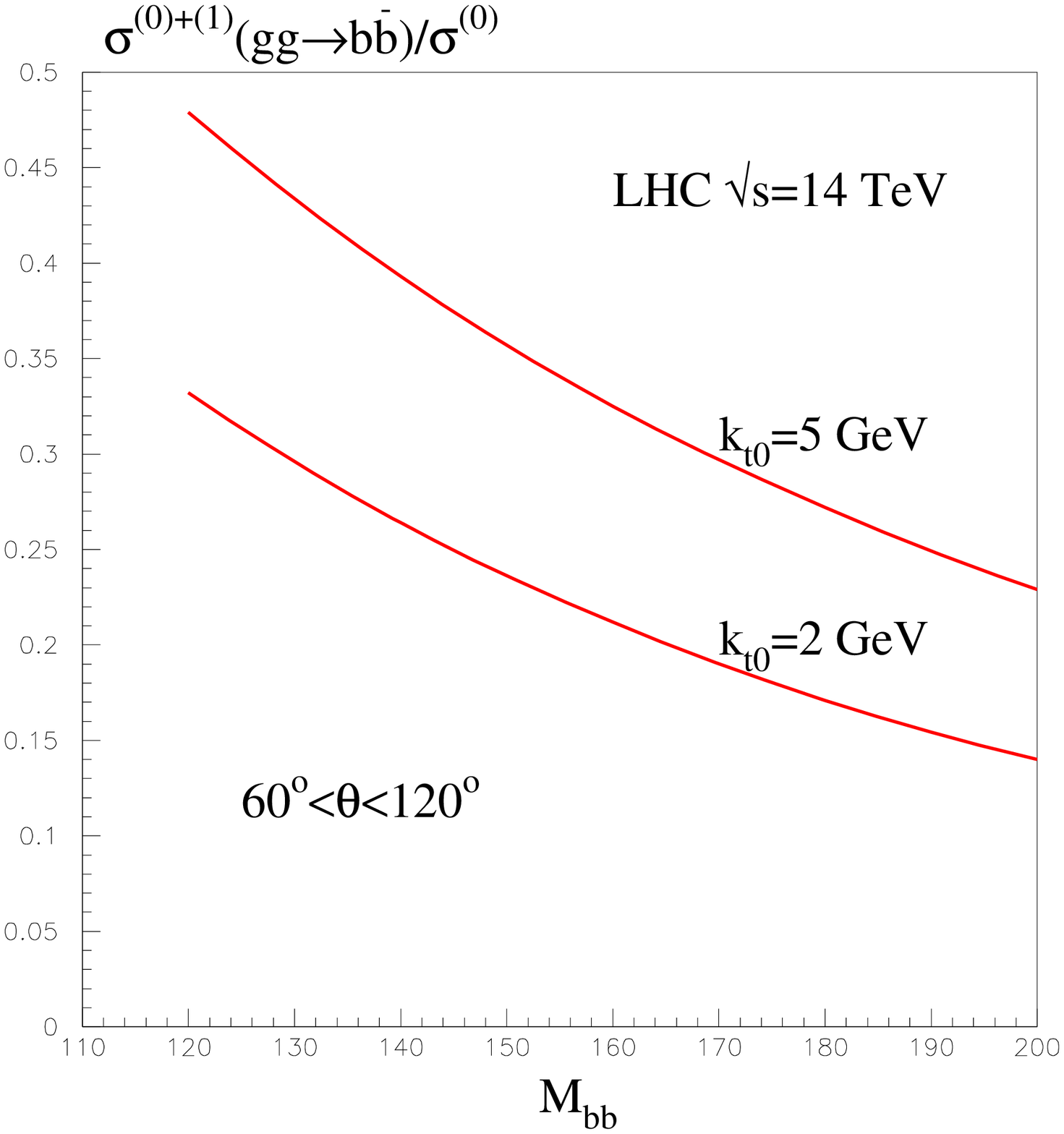}
\caption{\sf The mass dependence of the ration of the NLO exclusive
 $b\bar b$ cross section to that calculated in Born approximation.}
 \label{fig:ggbm}
\end{center}
\end{figure}

Finally, in Fig.~\ref{fig:ggbm}, we present the analogous ratio,
$\sigma^{(0)+(1)}/\sigma^{(0)}$, of the CED cross sections, integrated over
the region of $|\cos\theta|<1/2$,  expected at the LHC for exclusive
$b\bar b$ production in the central region (with the rapidity of the
$b\bar b$-pair $y=0$). In this case the
$gg^{PP}\to b\bar b$ amplitudes $T^{\rm Born}_1$ and $T_1$, which enter the
cross section (\ref{eq:cros}), were convoluted with the luminosity
amplitude
(the integrand of (\ref{eq:lum})) following the
$Q_t$ factorisation prescription; that is the amplitude $T_1(Q_t)$ was
included inside the $Q_t$ integral in (\ref{eq:lum}).
Again we show the results for two values of the infrared cutoff in the $b$-jet
definition -- $k_{t0}=5$ GeV and $k_{t0}=2$ GeV -- as the function of the
mass $M_{bb}$ of the $b\bar b$ pair for the scale equal to
$M_{bb}/2$.
Of course, for the smaller value, $k_{t0}=2$ GeV, we have a stronger suppression, but
this does not mean that by selecting narrower $b$-jets (with a smaller
cone size $\Delta R$ or a smaller $k_{t0})$ we can improve the Higgs signal-to-background ratio. The signal is diminished in the same way as the background when we
suppress the emission of an additional gluon in the $H\to b\bar b$ decay;
both the exclusive $b\bar b$ cross section and the  $H\to b\bar b$ signal
have the same Sudakov suppression.
Thus, in order not to lose  statistics, it is better not to take the value of $k_{t0}$ to be too small. On the other hand, $k_{t0}$ should not be too large.
Otherwise, we will not sufficiently suppress the three-particle, $gg^{PP}\to b\bar b+g$,
background \cite{krs}. From this viewpoint the predictions corresponding to $k_{t0}=5$ GeV
look appropriate.

In conclusion, the good news is that the radiative
QCD (one-loop) corrections {\it suppress the exclusive
$b\bar b$ background} (by a factor about  2,
or more for larger $b\bar b$-masses) for central exclusive diffractive (CED)
Higgs production,
in comparison with that calculated using the Born $gg^{PP}\to b\bar b$
amplitude. As discussed in \cite{KMRJz,DKMOR}, $b\bar b$
production in the $|J_z|=2$ state is
another background, which cannot be
distinguished from the $H\to b\bar b$ decay. However, this contribution can be
suppressed by selecting events with smaller transverse momenta of
the forward outgoing protons \cite{KMRJz}. Therefore, the exclusive gluon-gluon dijet production becomes the most
important background for the CED Higgs process, see \cite{DKMOR, hkrstw} for a detailed discussion. In order to
suppress further this QCD background we need better experimental discrimination between
$b$-quark and gluon jets; that is, to achieve a lower probability
$P_{g/b}$ for misidentifying a gluon as a $b$-jet.

\section*{Acknowledgements}
We thank Nigel Glover, Kemal Ozeren, Sven Moch, James Stirling, Robert Thorne, and especially
Adrian Signer, for useful discussions.
 MGR thanks the IPPP at the University of
Durham for hospitality, and VAK is grateful to the Kavli Institute for Theoretical Physics for hospitality during the completion of this work.
The work was supported by INTAS grant 05-103-7515, by grant RFBR 07-02-00023 and by the Russian State grant RSGSS-5788.2006.02.

 \section*{Appendix A:  Master integrals}

We present below the scalar master integrals needed to compute the one-loop effects shown in Fig.~\ref{fig:ggqq}. The results contain the functions
$$
F(\xi)\,\equiv\,\int_0^\xi dx\,\frac{\ln(1+x)}{x}\,=\,-Li_2(-\xi),
$$
$$
\rho\,\equiv\,\sqrt{s(s-4m^2)},
$$
$$
C_\epsilon\,\equiv \, -\frac 1\epsilon\,-\,\ln\frac{m^2}{4\pi \mu^2},
$$
where we work in $D=4+2\epsilon$ dimensions.

\subsection*{Extra gluon between quark lines}
Here, we give the integrals for diagrams 1-4 of Fig.~\ref{fig:ggqq}, which involve the virtualities
$$
a\equiv k^2-m^2,~~~  b\equiv (k-p_B)^2-m^2,~~~   c\equiv (k+p_A-p_1)^2-m_g^2,~~~  d\equiv (k+p_A)^2-m^2.
$$
$$
I^q_{0000}(s,t)~\equiv ~\int \frac{d^4k}{i\pi^2}\,\frac
1{abcd}\,=
$$
$$
=\,\frac 1{m^2-t}\,
\frac1{\rho}\,2\,\left[\,\ln\frac{\rho-s}{\rho+s}\,
\left(\ln\biggl(1-\frac t{m^2}\biggr)\,-\,
\frac 12\ln\frac{m_g^2}{m^2}\right)\,+\,
F\biggl(\frac s\rho\biggr)\,-\,F\biggl(-\frac s\rho\biggr)\right],
$$
$$
I^q_{0001}(t)~\equiv ~\int \frac{d^4k}{i\pi^2}\,\frac
1{abc}\,=\frac1{m^2-t}\,
\left[\,F\biggl(-\frac{t}{m^2}\biggr)+\frac{\pi^2}{6}\right],
$$
$$
I^q_{0010}(s)~\equiv ~\int \frac{d^4k}{i\pi^2}\,\frac
1{abd}\,=\,\frac 1{2s}\,
\ln^2\biggl(-\frac{s+\rho}{s-\rho}\biggr),
$$

%$$
%I^q_{1000}(s)~\equiv ~\int \frac{d^4k}{i\pi^2}\,\frac
%1{bcd}\,=\,-\frac1{\rho}\,
%\left[\,\ln\frac{m^2}{m_g^2}\,\ln\frac{\rho-s}{\rho+s}
%\,+\,2F\biggl(-\frac{\rho- s}{2\rho}\biggr)\, \right.
%$$
$$
\,\left. -\,2F\biggl(-\frac{\rho+ s}{2\rho}\biggr)-\,F\biggl(-\frac
{s+\rho}{s-\rho}\biggr)\,+\,F\biggl(-\frac{\rho-s}{s+\rho}\biggr)\right],
$$
$$
I^q_{1010}(s)+C_\epsilon~\equiv ~\int \frac{d^4k}{i\pi^2}\,\frac1{bd}\,=\,
\left[\frac{\rho}{s}\,\ln\biggl(-\frac{s-\rho}{s+\rho}\biggr)+2\right]+C_\epsilon,
$$
$$
I^q_{0101}(t)+C_\epsilon~\equiv ~\int \frac{d^4k}{i\pi^2}\,\frac1{ac}\,=\,
\left[2\,-\,\biggl(1-\frac{m^2}{t}\biggr)\ln\biggl(1-\frac{t}{m^2}\biggr)\right]+C_\epsilon,
$$
$$
\int \frac{d^4k}{i\pi^2}\,\frac1{cd}\,=\int \frac{d^4k}{i\pi^2}\,\frac1{bc}\,=~~C_\epsilon+2,
$$
$$
\int \frac{d^4k}{i\pi^2}\,\frac1{ad}\,=\int \frac{d^4k}{i\pi^2}\,\frac1{ab}\,=~~C_\epsilon.
$$

\subsection*{Extra gluon coupling to gluons}
Here, we give the integrals for diagrams 5 and 6 of Fig.~\ref{fig:ggqq}, which involve the virtualities
$$
a'\equiv k^2-m_g^2,~~~  b'\equiv (k-p_B)^2-m_g^2,~~~   c'\equiv (k+p_A-p_1)^2-m^2,~~~  d'\equiv (k+p_A)^2-m_g^2.
$$
$$
I^g_{0000}(s,t)~\equiv ~\int \frac{d^4k}{i\pi^2}\,\frac1{a'b'c'd'}\,=\,-\,\frac 1{s}\,\frac
1{m^2-t}\,\left[2\,\ln\frac{-s}{m_g^2}\,\ln\frac{m^2-t}{m\,m_g}\,
-\,\frac{\,\pi^2}{2}\right],
$$
$$
I^g_{1000}(s)~\equiv ~\int \frac{d^4k}{i\pi^2}\,\frac1{b'c'd'}\,=\,\frac1{\rho}\,
\left[\,\ln\frac{-s}{m^2}\,\ln\biggl(-\frac{s+\rho}{s-\rho}\biggr)\,
+\,\frac
12\,\ln^2\!\biggl(-\frac{s+\rho}{s-\rho}\biggr)\,
\right.
$$
$$
+\,\ln^2\!\biggl(-\frac{2s}{\rho-s}\biggr)\,\left.-\,2\,F\biggl(-\frac{s+\rho}{s-\rho}\biggr)-2\,F\biggl(-\frac{s+\rho}{2s}\biggr)\,
-\,2\,F\biggl(\frac{2s}{\rho-s}\biggr)\,-\,\pi^2\,\right],
$$
$$
I^g_{0100}(t)~\equiv ~\int \frac{d^4k}{i\pi^2}\,\frac1{a'c'd'}\,=\,\frac1{t-m^2}\,
\left[\,F\biggl(\frac{t}{m^2-t}\biggr)\, +\,\frac
12\,\ln^2\biggl(\frac{m^2-t}{m^2}\biggr)\,\right.
$$
$$
-\,\ln\frac{m_g^2}{m^2}\,
\ln\biggl(\frac{m^2-t}{m^2}\biggr)+\,\left.\frac 14\,\ln^2\frac{m_g^2}{m^2}\,+\,\frac{\pi^2}{12}\,\right],
$$
$$
I^g_{1010}(s)+C_\epsilon~\equiv ~\int \frac{d^4k}{i\pi^2}\,\frac1{b'd'}\,=\,\left[2+\ln\frac{m^2}{-s}\right]+C_\epsilon,
$$
$$
I^g_{0101}(t)+C_\epsilon~\equiv ~\int \frac{d^4k}{i\pi^2}\,\frac1{a'c'}\,=\,
\left[2\,-\,\biggl(1-\frac{m^2}{t}\biggr)
\ln\biggl(1-\frac{t}{m^2}\biggr)\right]+C_\epsilon,
$$
$$
\int \frac{d^4k}{i\pi^2}\,\frac1{c'd'}\,=\int \frac{d^4k}{i\pi^2}\,\frac1{b'c'}\,=~~C_\epsilon+2,
$$
$$
\int \frac{d^4k}{i\pi^2}\,\frac1{a'd'}\,=~~C_\epsilon+ \ln\frac{m^2}{m_g^2}.
$$

\subsection*{Extra gluon between a quark and a gluon line}
Here, we give the integrals for diagrams 7-10 of Fig.~\ref{fig:ggqq}, which involve the virtualities
$$
\tilde{a}\equiv k^2-m_g^2, ~~~~\tilde{b}\equiv (k+p_B)^2-m_g^2,~~~~\tilde{c}\equiv (k+p_1-p_A)^2-m^2,~~~~\tilde{d}\equiv (k+p_1)^2-m^2,
$$
$$
\tilde{e}\equiv (k-p_A)^2-m_g^2, ~~~~\tilde{f}\equiv (k-p_2)^2-m^2,
$$
$$
I^{qg}_{0000}(u,t)\,=\,\int \frac{d^4k}{i\pi^2}\,\frac1{\tilde{a}\tilde{b}\tilde{c}\tilde{d}}\,=\,\frac 1{m^2-u}\,\frac
1{m^2-t}\,2\left[\,
\ln\frac{m^2-u}{m\,m_g}\,\ln\frac{m^2-t}{m\,m_g}\,
+\,\frac{\pi^2}{12} \right],
$$
$$
I^{qg}_{1000}(u)\,=\,\int \frac{d^4k}{i\pi^2}\,\frac1{\tilde{b}\tilde{c}\tilde{d}}\,=\,\frac1{u-m^2}\,
\left[F\biggl(-\frac{u}{m^2}\biggr)\,+\,\frac{\pi^2}{6}\right],
$$
$$
I^{qg}_{0010}(u)=\int \frac{d^4k}{i\pi^2}\,\frac1{\tilde{a}\tilde{e}\tilde{f}}=\frac 1{u-m^2}
\left[F\biggl(\frac{u}{\,m^2-u}\biggr)
+\frac 12\ln^{2}\frac{\,m^2-u}{m_g^2}\,
\right.
$$
$$
\left.- \,\frac 14\ln^2\frac{m^2}{m_g^2}\,+\,\frac{\pi^2}{12}\,\right],
$$
$$
I^{qg}_{1010}(u)+C_\epsilon~=~\int \frac{d^4k}{i\pi^2}\,\frac1{\tilde{e}\tilde{f}}~=\,\left[2\,-\,\biggl(1-\frac{m^2}{u}\biggr)
\ln\biggl(1-\frac{u}{m^2}\biggr)\right]+C_\epsilon,
$$
$$
\int \frac{d^4k}{i\pi^2}\,\frac1{\tilde{c}\tilde{f}}~=~C_\epsilon,~~~~~~~~\int \frac{d^4k}{i\pi^2}\,\frac1{\tilde{a}\tilde{e}}~=~C_\epsilon+ \ln\frac{m^2}{m_g^2},
$$
$$
\int \frac{d^4k}{i\pi^2}\,\frac1{\tilde{c}\tilde{e}}~=~\int \frac{d^4k}{i\pi^2}\,\frac1{\tilde{a}\tilde{f}}~=~C_\epsilon+2.
$$

\section*{Appendix B:  $\overline{\rm MS}$ renormalization}

It is useful to recall the well known one-loop renormalization
 in the $\overline{\rm MS}$-scheme.
The bare quark mass $m_0$ and coupling constant $\alpha_0$
are expressed through their renormalized  values $m$
and $\alpha_S$ in the renormalized amplitude $T^R$ . In addition,
the renormalization of the quark and gluon wavefunctions
has to be taken into account, which results in factors $Z_2$ and $Z_3$.
The relation between the bare and physical coupling
at one-loop order reads
$$
\alpha_0\,=\,\alpha\,\left[\,1\,+\,\frac 1{\epsilon}\,
\beta_{\,0}\,\alpha\,+\,
{\cal O}(\alpha^2)\,\right]\,\equiv \,
\alpha\,+\,\alpha_0^{(1)}(\alpha),
$$
$$
\beta_0\,=\,\frac{11}{3}N_c\,-\,\frac 23 n_f.
$$
%For adjoint quarks $\beta_0=7/3N_c$.
 Similarly
$$
Z_{2,3}\,=\,1\,+\,Z_{2,3}^{(1)},
$$
where $Z_{2,3}^{(1)}$ are given by the first-order
corrections to the quark and gluon propagators.
The correction to the bare quark mass,
$$
m_0(m)\,=\,m\,+\,m_0^{(1)}(m)\,=\,m\,+\,\Sigma(m),
$$
is determined by the first-order quark self energy $\Sigma(\slash p)$.
With these factors included, the first-order amplitude
takes the form
$$
T^R\,=\,T^{\rm one-loop}(\alpha_0,m_0)\,
+\,\bigl[Z_2^{(1)}+Z_3^{(1)}+m_0^{(1)}(m)
+\alpha_0^{(1)}(\alpha)\bigr]\,T^{\rm Born}.
$$

The quark one-loop self energy reads
$$
\Sigma(\hat p)\,=\,-\,\frac {\alpha_S}{4\pi}\,\Gamma\biggl(2-\frac D2\biggr)
\int_0^1dx\,\bigl[Dm\,+\,\overline x(2-D)\slash p\,\bigr]
\left(\frac{-x\overline x p^2+x m^2+\overline x m_g^2}{4\pi \mu^2 e^\gamma}
\right)^{\frac D2-2},
$$
where $\overline x \equiv 1-x$ and $\gamma$ is
Euler constant.  As a result
$$
\Sigma(m)\,=\,-\frac {\alpha_S}{4\pi}\,m\,C_F\,\bigl(3C_\epsilon+4\bigr)
+{\cal O}(\epsilon),
$$
$$
Z_2^{(1)}\,=\,-\left.\frac{\partial\Sigma}{\partial p}\right|_{p=m}\,=\,
-\frac
{\alpha_S}{4\pi}\,C_F\,\bigl(C_\epsilon+4-2\ln\frac{m^2}{m_g^2}\bigr)
+{\cal O}(\epsilon).
$$

The gluon polarization operator in Feynman gauge,
$$
\Pi^{\mu \nu}(p)\,=\,
\bigl(g^{\mu \nu}p^2-p^\mu p^\nu\bigr)\,\pi(p^2),
$$
reads to one-loop order
% $$
% \pi(p^2)\,=\,\frac{\alpha_S}{4\pi}N_c\,\frac{(3D-2)}{2(D-1)}
% \frac{\Gamma^2(\frac D2-1)}{\Gamma(D-2)}
% \Gamma(2-\frac D2)\left(-\frac{p^2}{4\pi \mu^2 e^\gamma}
% \right)^{\frac D2 -2},
% $$
$$
\pi(p^2)\,=\,\frac{\alpha_S}{4\pi}\,
\frac{\Gamma^2(\frac D2-1)}{\Gamma(D-2)}
\Gamma(2-\frac D2)\left(-\frac{p^2}{4\pi \mu^2 e^\gamma}
\right)^{\frac D2 -2}
\left[N_c\frac{3D-2}{2(D-1)}-n_f\frac{(D-2)}{D-1}\right].
$$
With $D=4+2\epsilon$, one gets
$$
Z_3^{(1)}\,=\,\pi(p^2)\,=\,\frac{\alpha_S}{4\pi}\,
\biggl[N_c\,\left(\frac 53 C_\epsilon+
\frac 53\ln\frac{m^2}{-p^2}+\frac {31}{9}\right)
\,\biggr.
$$
$$
-\biggl.\,\frac 12 n_f\,\left(\frac 43 C_\epsilon+\frac 43\ln\frac{m^2}{-p^2}
+ \frac{20}{9}\right)\,+\,{\cal O}(\epsilon)\biggr].
$$

%\bibitem{Way} A.De Roeck, V.A.Khoze, A.D.Martin, R.Orava, M.G.Ryskin,
%Eur. Phys. J. C25 (2002) 391.

\end{document}